\documentstyle[12pt]{article}

\def\hybrid{\topmargin -20pt  \oddsidemargin 0pt
      \headheight 0pt   \headsep 0pt
      \textwidth 6.25in 
      \textheight 9.5in 
      \marginparwidth .875in
      \parskip 5pt plus 1pt   \jot = 1.5ex}

\hybrid

\def\x{\times}
\def\ox{\otimes}
\def\o+{\oplus}
\def\ra{\rightarrow}
\def\lra{\longrightarrow}

\def\beqa{\begin{eqnarray}}
\def\eeqa{\end{eqnarray}}
                       
\newcommand{\un}{\underline}

\newcommand{\al}{\alpha}

\newcommand{\la}{\lambda}
\newcommand{\si}{\sigma}
\newcommand{\ga}{\gamma}

\newcommand{\C}{{\cal C}}
\newcommand{\D}{{\cal D}}
\newcommand{\E}{{\cal E}}

\newcommand{\M}{{\cal M}}

\newcommand{\cP}{{\cal P}}
\newcommand{\cS}{{\cal S}}
\newcommand{\cO}{{\cal O}}

\newcommand{\resetcounter}{\setcounter{equation}{0}}

\begin{document}
\thispagestyle{empty}
\rightline{LMU-ASC 17/12}
\vspace{2truecm}
\centerline{\bf \LARGE Complete Bundle Moduli Reduction}
\vspace{.3truecm}
\centerline{\bf \LARGE in Heterotic String Compactifications}

\vspace{1.5truecm}
\centerline{Gottfried Curio\footnote{gottfried.curio@physik.uni-muenchen.de; 
supported by DFG grant CU 191/1-1}} 

\vspace{.6truecm}

\centerline{{\em Arnold-Sommerfeld-Center 
for Theoretical Physics}}
\centerline{{\em Department f\"ur Physik, 
Ludwig-Maximilians-Universit\"at M\"unchen}}
\centerline{{\em Theresienstr. 37, 80333 M\"unchen, Germany}}

\vspace{1.0truecm}

\begin{abstract}
A major problem in discussing heterotic string models
is the stabilisation of the many vector bundle moduli 
via the superpotential generated by world-sheet instantons.
In arXiv:1110.6315 we have discussed the method to make a discrete twist
in a large and much discussed class of vector bundles
such that the generation number gets new contributions
(which can be tuned suitably) and at the same time the space of bundle moduli
of the new, twisted bundle is a proper subspace (where the 'new', non-generic twist class exists)
of the original bundle moduli space;
one thus gets a model, closely related to the original model one started with,
but with enhanced flexibility in the generation number and where on the other 
hand the number of bundle moduli is {\em somewhat} reduced.
Whereas in the previous paper the emphasis was on examples for the new flexibility in the 
generation number we here classify and describe explicitly the twists and
give the precise reduction formula (for the number of moduli) for $SU(5)$ bundles leading to 
an $SU(5)$ GUT group in four dimensions.
Finally we give various examples where the bundle moduli space is reduced {\em completely}: 
the superpotential for such {\em rigid} bundles becomes a function of the 
complex structure moduli alone (besides the exponential Kahler moduli contribution).
\end{abstract}

\newpage

\section{Introduction}

We consider supersymmetric heterotic string models in four dimensions (4D) 
arising by compactification of the tendimensional theory on a Calabi-Yau
threefold $X$ endowed with a holomorphic vector bundle 
$V'=(V, V_{hid})$ with $V$ a stable bundle embedded 
in the visible $E_8$ whose commutant gives the unbroken gauge group in 4D.
We restrict our attention to $V$ and assume $c_1(V)=0$.

Besides the geometric (Kahler and complex structure) moduli of $X$ one gets 
moduli from the para\-meters of the bundle construction. As for the other 
moduli one wants to stabilise these moduli to particular 
values. This represents usually a formidable problem.

So it would be of interest to have a bundle construction which has no bundle moduli at all.
One way to reach such a rather unusual situation is to start with an ordinary bundle construction,
which comes with a bundle moduli space $\M_V$ of a large dimension
\beqa
\dim \M_V &=&h^1(End \, V)
\eeqa
and to make twists which are available only 
over a subset $\cS$ of the bundle moduli space $\M_V$
(cf.~[\ref{C2}] for a systematic introduction which also contains further references 
to related work): turning on such a twist
will restrict the moduli to $\cS$ if the twist is discrete. 
The availability of the twist poses a number $CON_{\gamma}$ of conditions
\beqa
\label{CON definition}
\sharp \{ \mbox{conditions for availability of the discrete twist $\gamma$} \}&=:&CON_{\gamma}
\eeqa
and reduces thereby the number for the new, twisted bundle $V'$ to
\beqa
\label{reduced moduli}
h^1(End \, V')&=&h^1(End \, V) - CON_{\gamma}
\eeqa
(in our spectral example the number of relevant $C$-deformations (cf.~below) is just reduced 
accordingly; no new, other moduli appear after the restriction to $\cS$).
If the twist really exists, as we assume, one can not, of course, put more 
conditions than free moduli available; that is we will assume always that
we are in a case $CON_{\gamma}\leq h^1(End \, V)$.

Clearly the goal in this line of thought is to make the number $h^1(End \, V')$ of
remaining moduli as small as possible. 
In the present paper we will give various examples where one can reach 
$CON_{\gamma}=h^1(End \, V)$, i.e.~{\em rigid} bundles, thus leaving only a zerodimensional space, 
i.e.~isolated values for all the bundle moduli 
(it might be a single point of the original space $\M_V$).
To avoid misunderstandings we point out that this is not ordinary moduli stabilisation:
we do not stabilise the bundle moduli of the original bundle $V$; rather we switch to a new,
closely related bundle $V'$ whose numerical, i.e.~cohomological characteristics are however
rather 'near' to the original bundle (and which has even more 'tuning freedom' from 
new discrete parameters)
and whose bundle moduli space is a zerodimensional subspace of the original space $\M_V$.

It is interesting to note that the issue of this type of moduli reduction 
converges also with another line of research in the heterotic context: 
in [\ref{cxstrfix}] heterotic constructions are made which exist only for a subset of the 
complex structure moduli space of an original model, 
leading to a corresponding reduction of freedom in that moduli space.

{\em Structure of the paper}

In {\em sect.~\ref{Set-up}} we recall the set-up [\ref{C2}] of the non-generic twists in the 
spectral bndle construction of the heterotic string. In {\em sect.~\ref{Twist classes}} we 
given an overview over the twist classes and describe them explicitly.
In {\em sect.~\ref{Main example}} we choose the case of a 4D GUT group $SU(5)$ which
corresponds to an $SU(5)$ vector bundle and determine $CON_{\gamma}$ for the main example
(out of two possibilities) for $\ga$ in this case (the more technical proof of the part (iii)
of proposition 2 is postponed to the appendix); finally we give a list of examples of rigid bundles 
and conclude with an outlook on some perspectives for further research 
in {\em sect.~\ref{Conclusions}}.

\section{\label{Set-up}A concrete set-up}

\resetcounter

We consider spectral $SU(n)$ vector bundles on an elliptic Calabi-Yau space
$\pi\!\!:\!X\ra B$ with section\footnote{We will identify notationally $\sigma$, its image and 
the divisor and cohomology class of that image; we also use the notation $c_1:=c_1(B)$, 
often with the pull-back to $X$ or $C$ understood, cf.~[\ref{FMW}].}
$\sigma$. In this case one has $V=p_*(\cP\ox p_C^* L)$
where (for this standard construction cf.~[\ref{FMW}])
one chooses a (ramified) $n$-fold cover surface $C\subset X$ over $B$, 
of cohomology
class $n\si + \pi^* \eta$ with\footnote{\label{base-point footnote}There are further conditions:
note first that the effectiveness of $C$ entails the effectiveness of $\eta$; furthermore the 
irreducibility of $C$ (which one needs to assume for the stability of $V$) is given just
for $\eta-nc_1$ effective and the linear system $|\eta|$ being base-point free;
the latter condition holds on a Hirzebruch surface ${\bf F_k}$ 
if $\eta \cdot b\geq 0$ and on a del Pezzo surface ${\bf dP_k}$ 
with $2\leq k \leq 7$ if $\eta \cdot E\geq 0$ for all curves $E$ with 
$E^2=-1$ and $E\cdot c_1=1$ (such curves generate the effective cone) 
(for notation cf.~the end of sect.~\ref{reduction counting}).}
$\eta\in H^{1,1}(B)$, and a line bundle $L$ over $C$; 
$\cP$ is the Poincare bundle over $X_{(1)}\x_B X_{(2)}$ restricted here to $X\x_B C$ and $p$ and $p_C$
the projections to the first and second factor, respectively.

The condition $c_1(V)\!=0$ will fix $c_1(L)$ in $H^{1,1}(C)\cap H^2(C, {\bf Z})$
up to a class $\ga$ in $ker (\pi_{C*})$: one has $c_1(L) \!=\! \frac{n\si + \eta + c_1}{2}+\ga$
where one has $\pi_{C*}\ga\!=\!0$ (here $\pi_C\!:\! C \!\ra\! B$ is the restricted projection; 
we will usually suppress the pull-back notation 
and write just $\phi$ for $\pi^* \phi$ or $\pi_C^* \phi$).
If one assumes that $C$ is ample one has $H^{1,0}(C)\!=\!0$ and $L$ is 
determined by its first Chern class (no further continuous moduli occur);
then also the curve $A_B\!:=\!C\!\cap\! B\!\subset\! B$ 
of class $\bar{\eta}:=\eta-nc_1$ is ample. 
The equation for $C$ is given by $w=a_0 z^2 +a_2 xz + a_3 yz + a_4 x^2 +a_5 xy=0$
for the cases (which are also phenomenologically the most important ones)
$n=4$ or $5$, resp.~(with $a_5=0$ for $n=4$; here $x,y,z$ are Weierstrass coordinates 
of the elliptic fibre and $a_i$ are 
global sections of\footnote{with the common abuse of notation (on our rational base surface $B$)
to denote divisor classes by symbols for corresponding cohomology classes} 
$\cO_B(\eta-ic_1)$. 
One can consider also the $\tau$ action $y\lra -y$ 
which gives the inverse in the group law on the fibers.

If one wants to describe the possible freedom one has in choosing $\ga$, 
generically one can say only the following: the only obvious classes on $C$ are, 
besides the section $\si|_C$, 
the pull-back classes $\pi^* \phi$ where the class $\phi$ comes from the base. 
One finds [\ref{FMW}] that 
$\pi_{C*} \si|_C = \bar{\eta}:=\eta - n c_1$ and so the only class in
$ker (\pi_{C*})$ available in general is
\beqa
\label{standard ansatz}
\ga &=& n\si|_C - \pi_C^* \bar{\eta}
\eeqa
(or suitable multiples $\la \ga$ of it; at this point 
an integrality issue occurs
which we do not make explicit here; important is that $\la$ has only discrete freedom). 
For the (negative of the) generation number $N_{gen}$ 
one gets $\frac{1}{2}c_3(V) = \la \eta \bar{\eta}$ (cf.~[\ref{C}]).

Now let us assume that, at least for a certain subset ${\cal S}$ of the moduli
space $\M_V$, further divisor classes on $C$ exist
(such that further corresponding cohomology classes, 
denoted by $\tilde{\chi}$ below, in the expression for $\ga$ can occur). 
Then we can make a more general ansatz for the cohomology class $\ga$ 
(where $\rho$ here is still a class coming from the base)\footnote{the pull-back operation itself 
is suppressed, so $\rho$ is actually $\pi_C^* \rho$; if no confusion can arise we will also 
suppress in the following the restriction and write just $\si$ for the class $\si|_C$} 
\beqa
\label{preliminary gamma ansatz}
\ga &=& n\si + \rho +\tilde{\chi}
\eeqa
The condition $\pi_{C*}\ga=0$ amounts now to 
$n(\bar{\eta}+\rho)+\pi_{C*}\tilde{\chi}=0$; 
to secure the divisibility of $\pi_{C*}\tilde{\chi}$ by $n$ 
we are led to the slightly modified ansatz $\tilde{\chi}:=n\chi$, 
that is
\beqa
\label{gamma ansatz}
\ga &=& n (\chi + \si) + \rho = n(\chi + \si)- \pi_{C*}(\chi+\si)
\eeqa
In the last rewriting we made manifest the condition on $\rho$
which guarantees\footnote{Note that the final term $\pi_{C*}(\chi+\si)$ is, in itself, 
a class projected down to $B$; 
if it occurs, as it is the case here, in a formula for the class $\ga$ on $C$,
then this means that it has to be read as being pulled-back to $C$, 
i.e.~this means actually the class $Q:=\pi_C^*\pi_{C*}(\chi+\si)$;
so both terms in the final expression $P-Q$ on the right hand side of 
(\ref{gamma ansatz}) fulfil $\pi_{C*}P=n\pi_{C*}(\chi+\si)=\pi_{C*}Q$, 
such that $\ga \in ker (\pi_{C*})$.}
$\ga \in ker (\pi_{C*})$
(again one may also consider suitable multiples $\la \ga$ and
an integrality issue occurs [\ref{C2}]).

If one turns on, as specified in (\ref{gamma ansatz}), 
a non-pull-back class $\tilde{\chi}=n\chi$ 
in the twist one gets (using $\pi_{C*}\si=\bar{\eta}=\eta-nc_1$; the formula is further evaluated
in (\ref{first new Ngen}), cf.~also [\ref{C2}]) 
\beqa
\label{Ngen contribution}
-N_{gen}&=&
\la\Big[ \eta \bar{\eta} + \pi_{C*}\chi\cdot \pi_{C*}\si - n\; \pi_{C*}(\chi\cdot \si)\Big]
\eeqa
(the first terms in the $[ ... ]$ brackets are the standard terms, the rest the corrections).

\section{\label{Twist classes}The non-generic twist classes}

\resetcounter

\subsection{\label{Lemma subsection}The two main example classes} 

In the following two examples of a non-generic twist class 
one uses the idea that under special conditions on the (bundle) moduli
one of the generically present classes $\pi_C^* \phi$ and $\si|_C$ can become reducible; 
then a component of this reducible class represents typically a 'new' class 
which can be used for a non-generic twist.

The {\bf first case} is the set-up where under certain conditions
the preimage $\C:=\pi_C^{-1}(c)$ of a curve $c\subset B$ becomes reducible in $C$
(a 'vertical' decomposition of a standard class)
\beqa
\label{vertical decomposition}
\mbox{{\em \un{vertical decomposition}}}\hspace{2cm}
\C=\pi_C^{-1}(c)&=&\C_1+\C_2\hspace{5cm}
\eeqa
We add a remark: whereas the full preimage $\C=\pi_C^{-1}(c)$ 
is an $n$-fold cover of the base curve $c$, a component $\C_i$  
(which may of course itself be again reducible) is an $k_i$-fold cover of $c$, 
such that $n=k_1+k_2$ (we adopt the convention $k_1\geq k_2$). 
If none of the $k_i$ is equal to $1$ we say we are in the 
case of an {\em ordinary} vertical decomposition, while if one of the $k_i$ is equal to $1$ 
we speak of a {\em special} vertical decomposition:
\beqa
\mbox{{\em ordinary vertical decomposition}}\hspace{2cm}
k_1\neq 1 \;\; \mbox{and} \;\; k_2 \neq 1 \hspace{3.6cm}\\
\mbox{{\em special vertical decomposition}}\hspace{2.4cm}
k_1=1 \;\;\; \mbox{or} \;\;\; k_2=1 \hspace{3.6cm}
\eeqa
(Ordinary vertical decompositions exist for $n\geq 4$.)
We note that one will always have the special vertical decomposition related to 
$c=A_B=C\cap B\subset B$ 
which can also be considered as $A_C:=B\cap C = \si|_C\subset C$ 
and to lie in $C$ and thus gives itself a component $\C_2$ of 
$\C=\pi_C^{-1}(c)$; in this case, however, the other component $\C_1$ 
in this $(n-1)+1$ split will not, of course,
represent a new class as one has $\C_1=\pi_C^*c-\si|_C$. We call this the {\em trivial}
special vertical decomposition.

In the {\bf second case} the curve $\si|_C=A_C$ becomes reducible in $C$ (or equivalently in $B$)
under certain conditions (a 'horizontal' decomposition) 
\beqa
\label{horizontal decomposition}
\mbox{{\em \un{horizontal decomposition}}}\hspace{2cm}
\si|_C&=&\D_1+\D_2\hspace{5.5cm}
\eeqa

One has the following injective association relation between these different new classes
\beqa
\{\mbox{{\em horizontal decomposition classes}}\}\!&\hookrightarrow & \!
\{\mbox{{\em nontrivial special vertical decompositions}}\}\;\;\;\;\;\;\;
\eeqa
(under $\D_i\lra c:=\D_i$).
To see this note that if one takes $\D_i$
as\footnote{recall that $\si|_C=A_C=A_B=B\cap C$ can be considered at the same time 
to lie in $C$ and in $B$} 
$c\subset B$ the preimage $\C$ decomposes always as a $(n-1)+1$ split because $\D_i=c$
lies (as $\C_2$) also in the surface $C$. 

Note further that on a fibre $F_b$ over $b\in c$ the $n$ points $q_i$ of $\C$ 
satisfy $\sum_{i=1}^n q_i=0$ in the gruop law of the elliptic fibre. 
By contrast one does not necessarily have, in the case of a vertical decomposition, 
$\sum_{i=1}^{k_1} q_i=0$ (or equivalently $\sum_{i=k_1+1}^n q_i=0$). 
If this happens in a vertical decomposition we will call it spectral
\beqa
\mbox{{\em spectral vertical decomposition}}\hspace{2.4cm}
\sum_{i=1}^{k_1} q_i=0 \hspace{3.6cm}
\eeqa
We will also call the class $\C_1$ (and equally then $\C_2$) spectral. One gets then
the following $1:1$ association relation (as being spectral for a $1$-cover means just to lie in $B$)
\beqa
\{\mbox{{\em horizontal decomposition classes}}\}\!\!\stackrel{1:1}{\leftrightarrow} \!\!
\{\mbox{{\em spectral nontrivial special vertical decompositions}}\}\,
\eeqa

Clearly it is desirable to have a criterion in term of a new class $\chi$ alone 
which allows to decide whether it is spectral (i.e. whether its fibre points sum to zero).
Here one has the following 
(note that even if $\chi$ is $\tau$-invariant the full preimage $\pi_C^*\pi_{C*}\chi$ 
is usually not, not to speak of the total spectral surface $C$; 
note also that $k_2=1$ or $2$ for $3\leq n \leq 5$)

\noindent
{\bf \un{Lemma}} {\em Let $\chi=\C_1$ or $\C_2$ be a 'new' class from a vertical decomposition.\\
(i) \hspace{.1cm}Let $3\leq n$:  
\hspace{1cm}$\chi$ is $\tau$-invariant $\Longrightarrow$ $\chi$ is spectral.\\
(ii) Let $3\leq n \leq 5$: $\C_2$ is $\tau$-invariant $\Longleftrightarrow$ $\chi$ is spectral.}

\subsection{The effective twist classes}

Clearly in both cases the new (i.e.~not generically available) divisor class
is given by an effective curve. 
Let us consider more generally the case where a given new divisor class contains an effective curve;
we will call such a class an {\em effective twist class}. So one has
\beqa
\{\mbox{{\em vertical decomposition classes}}\}&\subset &
\{\mbox{{\em effective non-generic twist classes}}\}\;\;\;\;
\eeqa
Let us ask whether the examples provided by the vertical decompositions classes 
exhaust already the effective non-generic twist classes.
So let $h^0(C, \cO_C(\chi))>0$ and assume we have in the expression for this cohomological number
(which effectively depends only on the linear equivalence class of the divisor)
a general member of this class already exchanged by an effective member; in other words
$\chi$ should denote already an effective divisor on $C$. 
Consider then $\C\!\!:=\pi_C^{-1}(\pi_{C}(\chi))$:
this is an ordinary pullback class $\pi_C^{-1}(c)$ (with $c=\pi_{C}(\chi)$) which furthermore
set-theoretically contains $\chi$; as $\chi$ is assumed to be a new class, it must be a proper
subvariety of this pullback curve; but the assumed existence of this subvariety means that 
$\chi$ is a component of the (then reducible) curve $\C$. 
\beqa
\label{eff twist classes}
\{\mbox{{\em vertical decomposition classes}}\}&=&
\{\mbox{{\em effective non-generic twist classes}}\}\;\;\;\;
\eeqa
We will adopt the terminology 'special/ordinary' also for the effective twist classes.

\section{\label{Main example}The spectral ordinary effective twist classes 
for $n\!\! =\!\! 5$}

\resetcounter

We will focus now on the case $n=5$ and consider certain non-generic twist classes
which are effective (non-effective ones are, of course, just differences of the effective ones).
We concentrate here on the ordinary classes which are spectral.
In sect.~\ref{Characterization} we characterize this abstractly defined type of new classes 
by an explicit construction and give a precise condition 
for the occurrence of this type of classes; in sect.~\ref{reduction counting} 
we compute the number of conditions relevant here and specify precisely the assumptions
one needs to make these computations (the most technical part is postpoend to the appendix); 
in sect.~\ref{rigid} we give some cases where
the number of necessary conditions is so large that all bundle moduli are already frozen
by choosing the corresponding 'new' (i.e.~non-generic) twist class.

\subsection{\label{Characterization}Characterization of the new classes}

Consider in a first, {\em preliminary} step the following factorisation
\beqa
\label{global factorization}
w&=&(f_1 z + g_1 x + h_1 y) \, (f_2 z + g_2 x) 
\eeqa
of the spectral cover equation
(here $f_1, g_1, h_1, f_2, g_2$ are sections of suitable line bundles over $B$).
If the coefficients $a_i$ can be written in this special way one gets 
(from the ensuing expressions for the $a_i$ like $a_0=f_1f_2, a_2=f_1g_2+f_2g_1$ and so on)
the relation 
\beqa
\label{factorization relation}
Res\; := \; a_0a_5^2-a_2a_3a_5+a_3^2a_4&=&0
\eeqa
(this means here identical vanishing over all of $B$, not yet an equation for
a curve in $B$).

Now we do in a second step the {\em proper} construction:
let $c$ denote a smooth irreducible reduced curve in $B$ and assume, 
instead of the global factorisability, only the following factorisability 
of $w$ over the elliptic surface $\E_c := \pi^{-1}(c)$
(with $F_1  := f_1|_c$ and so on)\footnote{where $F_1, G_1, H_1, F_2, G_2$ 
are now sections of suitable line bundles over $c$: 
for example one has that $F_1\in H^0\Big(c, \cO_c\Big((\eta-2c_1)|_c-(G_2)\Big)\Big)$
as $A_2=F_1\, G_2 + F_2\, G_1$ (with $A_i:=a_i|_c$) and so on}
\beqa
\label{factorization over c}
w|_{\E_c}&=&(F_1 z + G_1 x + H_1 y) \, (F_2 z + G_2 x) \, |_{\E_c}
\eeqa
If condition (\ref{factorization over c}) is fulfilled 
one has the ordinary vertical decomposition (\ref{vertical decomposition})
with $\C_1$ and $\C_2$ corresponding to the first and second factors 
in (\ref{factorization over c}), respectively: the five-fold cover $\C$ of $c$
decomposes into a triple cover $\C_1$ and a double cover $\C_2$; the individual factors 
have itself again the form of a spectral cover polynomial for $n=3$ and $2$, respectively,
so the decomposition is spectral (i.e.~the points in the individual $\C_i$ sum to zero fibrewise).
Note also that conversely, if one has such a component of $\C$ which is spectral
and has {\em good projection} (i.e.~$c=\pi_C(\chi)$ is smooth irreducible reduced)
then there will exist a corresponding spectral cover curve equation which must be a factor
in the sense of (\ref{factorization over c}).

So assume now conversely we have given an ordinary\footnote{assume we have, as we can in the argument, 
$\chi$ chosen already as an effective member of the divisor class}
effective non-generic twist class $\chi$ 
which is {\em spectral} and has {\em good projection};
we had seen in (\ref{eff twist classes}) that it actually comes from a vertical decomposition
over $c:=\pi_{C}(\chi)$ which in turn implies, as just pointed out, 
for spectral classes the factorization (\ref{factorization over c}). 
So one gets {\em in this case} the explicit description
\beqa
(\ref{vertical decomposition})\Longleftrightarrow (\ref{factorization over c}) 
\eeqa

For example, one gets\footnote{the first terms in the $[\, ...\, ]$ brackets 
on the right hand sides are the standard contributions, 
the terms proportional to $c$ are the new contributions} 
in this example by taking\footnote{taking $\chi= \C_1$ instead of $\C_2$ gives, 
with $(3\bar{\eta}-5(h_1))c$ as new term in (\ref{first new Ngen}), 
just the negative of the present new term
(as $\bar{\eta}\cdot c=A_B\cdot c=(a_5)\cdot c=(h_1)\cdot c+(g_2)\cdot c$)} 
$\chi=\C_2$ (as cohomology class) [\ref{C2}]
\beqa
\label{first new Ngen}
-N_{gen}\!\!\!&=&\!\!\!\la \Big[\eta \bar{\eta}+ \Big(2\bar{\eta}-5(g_2)\Big)c\Big]
\eeqa

Furthermore one gets as necessary condition for the factorizability over $c$ that the 
curve given in (\ref{factorization relation}) 
(now read as an equation 
for a curve in $B$ and not as an identical vanishing over all of $B$)
has $c$ as a component, 
i.e.~that the equation (\ref{factorization relation}) is fulfilled along $c$
\beqa
\label{factorization relation over c}
Res|_c\; = \; A_0A_5^2-A_2A_3A_5+A_3^2A_4&=&0
\eeqa
This gives the following logical implication
\beqa
(\ref{factorization over c}) \Longrightarrow (\ref{factorization relation over c})
\eeqa

Concerning the reverse arrow let us assume at first that $c\cong {\bf P^1}$: then 
the relation (\ref{factorization relation over c}) is not only necessary 
but also sufficient to have (\ref{factorization over c}). For this we show 
that if the relation (\ref{factorization relation over c}) holds one can introduce 
various polynomials $F_1, G_1, H_1, F_2, G_2$ on $c$ which turn out to have
the necessary relations to the $A_i$ to get (\ref{factorization over c}), i.e.~
\beqa
A_0z^2+A_2xz+A_3xy+A_4x^2+A_5xy&=&(F_1 z + G_1 x + H_1 y) \, (F_2 z + G_2 x) 
\eeqa
So the relations one needs to derive are
\beqa
A_0&=&F_1F_2\\
A_2&=&F_1G_2+G_1F_2\\
A_3&=&H_1F_2\\
A_4&=&G_1G_2\\
A_5&=&H_1G_2
\eeqa
Now note first that because of (\ref{factorization relation over c}) one has $A_5|A_3A_4$,
so that one can write $A_5=H_1 G_2$ with $H_1|A_3$ and $G_2|A_4$ for some polynomials $H_1, G_2$; 
let us write furthermore $A_3=H_1F_2$ and $A_4=G_1G_2$ with certain polynomials $F_2, G_1$.
From (\ref{factorization relation over c}) one gets $A_3G_2|A_0A_5$ 
such that also $A_5F_2|A_0A_5$ (as $A_5F_2=A_3G_2$), 
such that $F_2|A_0$ and one can write $A_0=F_1F_2$ for some $F_1$. 
From these determinations it follows already,
once more with (\ref{factorization relation over c}), that\footnote{here $A_5\neq 0$ 
as otherwise $c\subset (a_5)=A_B$ and one would be 
in a special case (of a $4+1$ split instead of a $3+2$ split); 
above we did assume $A_3\neq0$: otherwise $A_0=0$ and one can take $F_2=0, G_2=1$} 
$A_2=(A_0A_5^2+A_3^2A_4)/A_3A_5=F_1G_2 + G_1F_2$. In other words one gets
\beqa
(\ref{factorization over c}) \stackrel{{\bf \;\; P^1}}{\Longleftarrow} 
(\ref{factorization relation over c})
\eeqa

Alternatively one might, to show this reverse implication, 
adopt the strategy of an argument of [\ref{FMW}].
This relies on a certain interpretation of the relation (\ref{factorization relation over c}) 
or, equivalently, $c\subset (Res)$ (cf.~(\ref{factorization relation})): 
the vanishing divisor $(Res)\subset B$ can be interpreted
as the locus of $b\in B$
where $\tau$-conjugated fibre points $q_i, q_j$ exist ($i\neq j$), i.e.~points with $q_i+q_j=0$ 
(note that $\tau$-conjugacy is just
$y\lra -y$ in coordinates such that $Res = Resultant(P, Q)$ is relevant where $w = P(x)+Q(x)y$).
So this is the locus in whose preimage in $C$ a {\em spectral} $2$-fold cover point set 
splits off the total $5$-fold spectral cover point set, so 
\beqa
c\subset (Res) & \Longleftrightarrow & \C=\C_1+\C_2 \; \mbox{with $\C_2$ spectral}
\eeqa
which shows in view of the Lemma in sect.~\ref{Lemma subsection} that 
(here $c$ is assumed to be smooth irreducible reduced, 
respectively $\chi$ is assumed to have good projection)
\beqa
\label{equivalence}
(\ref{factorization over c}) \; \Longleftrightarrow \;
\mbox{ordinary spectral effective vertical decomposition} \; \Longleftrightarrow \;
(\ref{factorization relation over c})
\eeqa

Let us summarize some points we have encountered.
If one restricts the bundle moduli (the degrees of freedom coming from the $a_i$) by posing 
the condition (\ref{factorization relation over c}) along $c$, 
one gets the factorization of the equation (\ref{factorization over c}) for $\C$ 
and thus the decomposition (\ref{vertical decomposition}) 
which defines the 'new' class of $\C_1$ (or equivalently the 'new' class $\C_2$).
Asking conversely which moduli restriction is enforced
by demanding the existence of this class (because it is used in 
a discrete twist) one has to take into account two things: first, what one really
uses in the twist construction is a line bundle, thus a divisor {\em class} on $C$; so,
for the purpose of our procedure,
one has to make sure that an {\em effective} representative in this class exists;
this is the reason we have assumed from the outset 
that we treat the case of an effective non-generic twist class.
One also has to clarify whether the existence of such a curve
(which we hope to play the role of $\C_1$)
can arise {\em only} in the way (\ref{vertical decomposition}), resp.~(\ref{factorization over c}),
or whether it may exist 'accidentally' already on a larger moduli subspace 
than the one given by (\ref{factorization relation over c})
(where it exists 'naturally'); this issue is treated and solved in (\ref{equivalence}).

\subsection{\label{reduction counting}The question of moduli reduction}

We now want to count the precise number of the conditions which describe the circumstances
in which the type of 'new' class considered in (\ref{equivalence}) exists, 
cf.~(\ref{CON definition}). We assume that $c$ is smooth irreducible and reduced 
(this is an assumption on $c$ if one starts from $c$; it restricts
the given twist class $\chi$ on $C$ with $c=\pi_C(\chi)$ to have good projection 
if one starts from $\chi$; the interesting possibilities of the case of $c$ being reducible 
will not be studied in the present paper, cf.~the discussion in the Outlook in sect.~\ref{Conclusions}).

We remark that $CON_{\gamma}$, where $\gamma=n\chi - \pi_{C*}\chi + n\si - \bar{\eta}$,
will be the same for $\chi = \C_1$ or $\C_2$ as their sum, the pullback class 
$\C:=\pi_C^{-1}(c)$, exists universally. We will therefore denote $CON_{\gamma}$ henceforth
simply by $CON_c$ (we have fixed that we are in the ordinary case, 
i.e.~in the case of the $3+2$ split of $n=5$).

Let us count now the number of conditions imposed on the $a_i$ by the factorization
(\ref{factorization over c}). As described above this implies
the vanishing (\ref{factorization relation over c}).
Consideration of the association $(a_i)\lra A_0A_5^2-A_2A_3A_5+A_3^2A_4$
gives a map $\M_V\lra |(3\eta-10c_1)|_c|$ 
or alternatively, before taking the overall scaling, 
\beqa
\label{the map}
\bigoplus_{i=0, \neq 1}^5 H^0\Big(B, \cO_B(\eta-ic_1)\Big)&\lra & 
H^0\Big(c, \cO_c\Big((3\eta-10c_1)|_c\Big)\Big)
\eeqa

We thereby get the following proposition 
for the number $CON_c$ of conditions posed inside $\M_V$ by demanding 
(\ref{factorization relation over c})

\noindent
{\bf \un{Proposition 1}}\\
{\em (i) One has the estimate 
\beqa
\label{CON estimate}
CON_c \;\; \leq \;\; h^0\Big(c, \cO_c((3\eta-10c_1)|_c)\Big) 
\;\; \leq \;\; \frac{1}{2}(6\bar{\eta}+11c_1-c)c
\eeqa
(ii) if the map (\ref{the map}) is surjective one has an equality in the first inequality in (i) \\
(iii) if the class $3\bar{\eta}+6c_1\!-\!c$ is effective one has
an equality in the second inequality in (i) \\
(iv) if the map (\ref{the map}) is surjective 
and the class $3\bar{\eta}+6c_1\!-\!c$ is effective one has}
\beqa
CON_c&=&\frac{1}{2}(6\bar{\eta}+11c_1-c)c
\eeqa
The assertions $(i),(ii)$ are obvious. To evaluate $h^0\Big(c, \cO_c((3\eta-10c_1)|_c)\Big)$ 
precisely we look 
for assumptions implying
$h^1\Big(c, \cO_c((3\eta-10c_1)|_c)\Big)\!\!=\!h^0\Big(c, \cO_c((c-c_1-(3\eta-10c_1))|_c)\Big)\!\!=\!0$.
Clearly it will be sufficient to assume that $(3\eta-9c_1-c)|_c$ is effective as a divisor on $c$.
We will just assume that $3\eta-9c_1-c=3\bar{\eta}+6c_1-c$ is effective as a divisor on $B$.
Then one gets $h^0\Big(c, \cO_c((3\eta-10c_1)|_c)\Big)=
(3\eta-10c_1)c+\frac{1}{2}e(c)=(3\bar{\eta}+5c_1)c+\frac{1}{2}(c_1-c)c$.

Let us now come back to point $(ii)$: 
one gets an equality in the first inequality in (\ref{CON estimate}) if the map (\ref{the map})
is surjective (otherwise one would overcount the real number of conditions on the $a_i$).
To make the proposition usefully applicable 
one has to find a criterion (implying the surjectivity in question) which can be checked easily.
Now the map in question can be factorised, in two different ways, as follows
\begin{center}
\begin{tabular}{ccccc}
$\bigoplus_{i=0, \neq 1}^5 H^0\Big(B, \cO_B(\eta-ic_1)\Big)$ & & $\stackrel{\alpha}{\lra}$ & & 
$H^0\Big(B, \cO_B(3\eta-10c_1)\Big)$ \\
 & & & & \\
$\downarrow$ & & & & $\downarrow\; \beta$ \\
 & & & & \\
$\bigoplus_{i=0, \neq 1}^5 H^0\Big(c, \cO_c((\eta-ic_1)|_c)\Big)$ & & $\lra$ & &
$H^0\Big(c, \cO_c((3\eta-10c_1)|_c)\Big)$
\end{tabular}
\end{center}
Here elements in the left upper space are mapped as follows
\begin{center}
\begin{tabular}{ccccc}
$\;\;\;\;\;\;\;\;\;\;\;\;\;\;\;\;\; (a_i)$ & $\;\;\;$ & $\lra$ & $\;\;\;$ 
& $a_0a_5^2-a_2a_3a_5+a_3^2a_4$ \\
 &  &  \\
$\;\;\;\;\;\;\;\;\;\;\;\;\;\;\;\;\;\downarrow$ & $\;\;\;$ & & $\;\;\;$ 
& $\downarrow$ \\
 & & & & \\
$\;\;\;\;\;\;\;\;\;\;\;\;\;\;\;\;\; (A_i)$ & $\;\;\;$ & $\lra$ & $\;\;\;$ 
& $A_0A_5^2-A_2A_3A_5+A_3^2A_4$ 
\end{tabular}
\end{center}
Now clearly it will be sufficient for the surjectivity of the total map if both factor maps
in a chosen factorization are surjective. 
From the two possible paths, the upper way (using as horizontal map the upper map)
and the lower way (using as horizontal map the lower map),
we will choose the upper way (the lower way is less suitable for our purposes 
as it leads to stronger conditions). 
One gets then the following 
(here assertion $(ii)$ follows as $H^1\Big(B, \cO_B(3\eta-10c_1-c)\Big)\!=\!0$ holds for 
$3\eta-9c_1-c\!=\!3\bar{\eta}+6c_1-c$ being ample; part $(iii)$ will be shown in the appendix)
\\
\noindent
{\bf \un{Proposition 2}} \\
{\em (i)The map (\ref{the map}), 
i.e.~the map $\beta \circ \alpha$, is surjective if the maps $\alpha$ and $\beta$
are surjective\\
(ii) the map $\beta$ is surjective if $3\bar{\eta}+6c_1-c$ is ample\\
(iii) the map $\alpha$ is surjective if $c_1$ and $\bar{\eta}$ are both big and nef.}\footnote{a
divisor $D$ is big if $D^2>0$ and nef if $DD'\geq 0$ for all effective divisors $D'$; an ample
divisor $D$ is in particular big and nef (the precise condition is being big and $DD'> 0$ 
for all effective divisors $D'\neq 0$)}

Note that in the standard examples\footnote{which are given by 
the Hirzebruch surfaces ${\bf F_k}$ ($k=0,1,2$), the del Pezzo surfaces ${\bf dP_k}$
($k=0, \dots, 8$) and the Enriques surface, cf.~the explanations given in the Remark below} 
for the rational base $B$ the class $c_1$ is always ample except for ${\bf F_2}$
(where it is however still big and nef) and the Enriques surface (where it is not even effective). 
The class $\bar{\eta}$ of the curve $A_B=\si\cap B\subset B$ is always effective.

Putting everything together one arrives at the following

\noindent
{\bf \un{Theorem}} {\em Assume that $3\bar{\eta}+6c_1-c$ is ample
and that $c_1$ and $\bar{\eta}$ are both big and nef: then one has}
\beqa
CON_c&=&\frac{1}{2}(6\bar{\eta}+11c_1-c)c
\eeqa

Note that if $3\bar{\eta}+6c_1-c$ is ample then the condition from proposition 1
that $3\bar{\eta}+6c_1-c$ is effective
is redundant as this class will then be 
such that sufficiently large multiples of it will be (cf.~footn.~\ref{ample footnote}) effective 
as will be then also the class itself
(in both cases up to linear equivalence what is sufficient for the use in proposition 1).

\noindent
\un{\em Remark:} For convenience of the reader we recall the essentials about the base surfaces
which constitute the standard examples (we will have no need to consider the Enriques surface).
The corresponding facts will be used below when we display some examples.

\noindent
\un{The Hirzebruch surfaces} 
The surface ${\bf F_k}$ is a ${\bf P^1}$-fibration over
a base ${\bf P_1}$ denoted by $b$ (the fibre is denoted by $f$;
as no confusion arises $b$ and $f$ will denote also the 
cohomology classes). One has $c_1({\bf F_k})=2b+(2+k)f$ and the curve
$b$ of $b^2=-k$ is a section of the fibration; 
there is another section ("at infinity") having the
cohomology class $b_{\infty}=b+kf$ and the self-intersection number $+k$; 
note that $b_{\infty}\cdot b = 0$.
A class 
\beqa
(x,y)&:=&xb+yf
\eeqa 
is ample exactly if$^{\ref{Hartshorne}}$ 
$(x,y)\cdot f > 0$ and $(x,y)\cdot b >0$, i.e.~if $x>0, y>kx$.
An irreducible non-singular curve of class $xb+yf$ exists exactly 
if\footnote{\label{Hartshorne}Cf.~Corollary 2.18, Chap.~V, 
{\em Algebraic Geometry}, R. Hartshorne, Springer Verlag (1977).} 
the class lies in the ample cone (generated by the ample classes) 
or is one of the elements $b, f$ or $ab_{\infty}$ (the last only for $k>0$; here $a>0$) 
on the boundary of the cone; these classes
together with their positive linear combinations span the effective cone 
($x,y\geq 0$). $c_1$ is ample for ${\bf F_0}$ and ${\bf F_1}$, whereas for 
${\bf F_2}$, where $c_1=2b_{\infty}$ (such that $c_1\cdot b=0$),
it lies on the boundary of the cone.

\noindent
\un{The del Pezzo surfaces} The surfaces ${\bf dP_k}$ are the blow-up of ${\bf P^2}$
at $k$ points $P_i$ for $k=0, \dots, 8$ 
(lying suitably general, i.e.~no three points lie on a line, no six on a conic);
the exceptional curves from these blow-ups are denoted by 
$E_i$, $i=1, \dots, k$ 
(one has ${\bf dP_1}\cong {\bf F_1}$ with $E_1$ corresponding to $b$).
The intersection matrix for $H^{1,1}({\bf dP_k})$ in the basis 
$(l, E_1, \dots, E_k)$, 
with $l$ the proper transform of the line $\tilde{l}$ from ${\bf P^2}$, 
is just $Diag(1, -1, \dots, -1)$; 
furthermore one has 
\beqa
c_1({\bf dP_k})&=&3l-\sum_i E_i
\eeqa 
such that $c_1^2({\bf dP_k})=9-k$.

\subsection{\label{rigid}An application: rigid bundles}

Finally we recall the number of vector bundle moduli 
(we assume that $C$, or equivalently $\bar{\eta}$, is ample, so no $\tau$-odd moduli occur). 
This is computed from the number of
different possible shapes of $C$ inside $X$, i.e.~by $h^0(X, \cO_X(C))-1$ which in turn
is computed either directly [\ref{het mod}] 
or from the corresponding numbers of degrees of freedom
in the coefficients $a_i$ of the spectral cover equation $w=0$ for $C$ [\ref{FMW}].
To get explicit results one assumes in both cases that $\bar{\eta}$ is ample (and $c_1$ also
ample or at least big and nef). One then gets
\beqa
h^1(End \, V)&=&
n-1+\Big(\frac{n^3-n}{6}+n\Big)c_1^2+\frac{n}{2}(\bar{\eta}+nc_1)\bar{\eta}+\bar{\eta}c_1
\eeqa
We get for the general equation $h^1(End \, V)=CON_c$ in our case
\beqa
\label{explicit rigidity condition}
8+50c_1^2+5\bar{\eta}^2+27\bar{\eta}c_1&=&(6\bar{\eta}+11c_1-c)c
\eeqa
Here are some solutions to this equation ($3\bar{\eta}+6c_1-c$ is always ample; 
recall $\eta=\bar{\eta}+5c_1$)
\begin{center}
\begin{tabular}{|c||c|c|c|c|c|c|c|}
\hline
$B$ & ${\bf F_0}$ & ${\bf F_1}$ & ${\bf F_2}$ & ${\bf P^2}$ & ${\bf dP_5}$ & ${\bf dP_7}$ & ${\bf dP_8}$\\
\hline
$\bar{\eta}$ & $(3,5)$  & $(3,7)$ & $(3,8)$ & $24l$ & $2c_1$ & $3c_1$ & $2c_1$\\
\hline
$c$ & $(15,21)$  & $(16,27)$ & $(15,36)$ & $38l$ & $9c_1$ & $9c_1$ & $11c_1$\\
\hline
\end{tabular}
\end{center}

The classes $\bar{\eta}$ and $c$ occurring here are actually ample, 
so on ${\bf F_k}$ and ${\bf P^2}$ (where $l$ denotes the class of the general line)
there exist smooth irreducible representative curves of the classes in question 
(for ${\bf dP_5}$ $c_1$ is known to be very ample so again smooth irreducible representatives
will exist by Bertini's theorem). 
In all cases $c_1$ is effective so the further conditions, cf.~footn.~\ref{base-point footnote},
amount just to $\bar{\eta}$ being effective (which obviously is the case) 
and the linear system $|\eta|$ being base-point free; the latter can be checked via the criteria
mentioned in footn.~\ref{base-point footnote} for the ${\bf F_k}$, for ${\bf P^2}$ (where it is 
obvious) and also for ${\bf dP_5}$ and ${\bf dP_7}$.

We stop with the checks performed on these special examples at this point although further issues
in these particular models could and should be clarified. In the context of the present paper 
these examples serve, however, merely an illustrative purpose: we wanted to show that one can fix 
circumstances (cf.~the assumptions in the Theorem of the previous subsection) in which one
can determine the number of conditions posed on the bundle moduli by the presence of specific
non-generic twists in a controlled manner (cf.~the propositions; this concerns essentially
prop.~1$(i)$ that under specific assumptions one can make sure that the reducibility of the preimage
$\C=\pi_C^{-1}(c)$ of a base curve $c$ poses actually the maximum number of conditions one expects) 
such that one gets a precise formula for $CON_{\ga}$; 
and secondly we wanted to show that there is no obstacle in principle
that this bundle moduli reduction is so effective that it becomes even complete.

\newpage

\section{\label{Conclusions}Conclusions and Outlook}

Supersymmetric particle physics models coming from heterotic string theory 
have, like other string models, to deal with a large number of 
occurring moduli [\ref{het mod}] and their potential stabilisation,
here concretely geometric (K\"ahler and complex structure) moduli and bundle moduli. 
As the stabilisation of the latter is a difficult and complex task
it is interesting to consider cases where only few or no bundle moduli occur. 
A possibility to achieve this is to make discrete modifications of a given bundle construction 
which are available only over a subset of the bundle moduli space 
such that the new, twisted bundle has less parametric freedom
(i.e.~turning on such discrete 'twists' constrains the moduli 
which thereby are restricted to a subset of their moduli space).

After occasional occurrence of this idea in the literature, 
usually in slightly different contexts like $F$-theory (cf.~for example [\ref{DW}]), 
this was considered systematically in the heterotic context in [\ref{C2}]. 
There the question was investigated in the most intensly studied class
of heterotic string models, that of spectral cover bundle constructions 
on elliptic Calabi-Yau spaces $X$ [\ref{FMW}]. General formulae were given 
for the influence of the twist on the cohomological data, 
most prominently the generation number $N_{gen}$, given by $c_3(V)/2$,
and various conrete examples were studied too.

In the present paper the emphasis is on the (bundle) moduli reduction effect of having turned on
a discrete twist. The 'new' divisor class on the spectral cover surface $C$ (from which the
class $\gamma$ is derived which corresponds to the line bundle twist) 
is here assumed to have an effective representative divisor $\chi$ 
such that geometrical methods can be brought to bear most directly. 
Such non-generic effective twists are described as arising from standard classes 
(like pullback classes $\pi_C^* c$, here with $c = \pi_{C*} \chi$) which become reducible.
After distinguishing and discussing various types of twist classes we focus on classes we call
'spectral' (where also for the component $\chi$, just as for a total pullback class $\pi_C^* c$, 
the points sum fibrewise to zero in the group law on the fibre); furthermore we assume that we are
in the 'ordinary' case, as we call it, where the corresponding decomposition of $\pi_C^* c$ 
with $c = \pi_{C*} \chi$ does not have a component lying itself already in the base (and thus
would have covering degree $1$ over $c$; 
such a case in general, even when the covering curve does not lie in the embedded base,
we call 'special'). These effective non-generic twist classes which are ordinary and spectral 
(and project smoothly to $B$)
can be described quite explicitly
by means of a factorization of the corresponding spectral cover equation (when restricted to the
elliptic surface $\E_c$ over $c$). The condition for such a factorization is provided,
in one of the phenomenologically most important cases which is given by $n=5$ (corresponding
to an $SU(5)$ GUT group in four dimensions),
by the relation (\ref{factorization relation over c}).

To compute the moduli reduction effect one thus has to count the number of 
conditions imposed by this relation.
This is straightforward in principle by computing the dimension of the space
$H^0\Big(c, \cO_c((3\eta-10c_1)|_c)\Big)$ in which the expression $A_0A_5^2-A_2A_3A_5+A_3^2A_4$ 
lives. There is, however, the problem to make sure that these expressions fill out that space
completely; afterall this element of the space $H^0\Big(c, \cO_c((3\eta-10c_1)|_c)\Big)$ 
has by definition built in two restrictions: first being composed in the indicated manner
from the expressions $A_i$, and secondly the latter themselves are restrictions from 
the corresponding global $a_i$ over $B$ to $c$; alternatively this restrictedness may be described
in the reversed order: $A_0A_5^2-A_2A_3A_5+A_3^2A_4$ is restricted from the corresponding global 
expression $a_0a_5^2-a_2a_3a_5+a_3^2a_4$ which furthermore itself is not a general element of the
space $H^0\Big(B, \cO_B(3\eta-10c_1)\Big)$ in which it lives but is an element composed of the 
elements $a_i$ in the indicated manner. What we have described here is the problem pointed out
in the diagrams before Proposition 2 in section \ref{reduction counting}. The solution to this
problem 
is given in Proposition 2; the proof of its most nontrivial part is given in the appendix. 

Armed with the precise formula for the reduction effect (most importantly with the precise
conditions under which the formula does hold which is here decisive 
due to the subtleties described above) 
one can search for examples of {\em complete} moduli reduction. 
We give in section \ref{rigid} various examples which satisfy the mentioned conditions.

\noindent
{\em Outlook}

Several things can now be done. 
First one should extend the counting method for the moduli reduction effect given here also to 
other spectral (effective) twist classes of phenomenological interest: 
i.e.~to the special classes for $n=5$ and to the ordinary and special classes for $n=4$ 
(corresponding to an $SO(10)$ GUT group in four dimensions).
Having the corresponding formulae for $CON_{\gamma}$ in these cases one may search for further rigid
bundles and also for a comprehensive approach to the computation of the number 
(\ref{reduced moduli}) of reduced moduli which produces these results beyond the case-by-case 
analysis.

Even more important conceptually, in connection with the topic of rigid bundles,
seem to be the following two points: first the treatment of the superpotential
and secondly the issue of making the described potential 'rigidification' of a bundle 
an effect as universal as possible.
Let us explain what we mean by these two lines of investigations.

First, having a rigid bundle, i.e.~a heterotic string model which lacks any bundle modu\-li,
the superpotential becomes now a 'function' of the complex structure moduli alone (besides the
well-known explicit exponential dependence on the Kahler moduli; their number for
$B$ a Hirzebruch surface or ${\bf P^2}$ is $3$ or $2$, respectively). As all of the difficult 
questions connected with describing the precise dependence of the world-sheet instanton generated 
superpotential on the vector bundle moduli collapse here dramatically
(as no continuous bundle moduli are left) one can hope to study now directly all sort of relevant
questions, like (Kahler and complex structure) moduli stabilisation 
(and value of the scalar potential at the stabilised point) and preservation or breaking
of supersymmetry, by making explicit the dependence on the remaining complex structure moduli 
(given in the coefficients $g_2$ and $g_3$ of the Weierstrass equation of the elliptic fibration). 
The {\em explicit} dependence of the superpotential on {\em these} 
geometric moduli has not been investigated very much (because it is usually intertwined 
with problems concerning the bundle moduli). Ideally one may think of an explicit
expression for the superpotential in terms of (the moduli contained in) $g_2$ and $g_3$ 
(together with the known Kahler moduli dependence), something still out of reach at present
in a nontrivial situation {\em with} vector bundle moduli (despite some progress for results
on individual Pfaffian summands of the full superpotential and modest steps towards the an 
understanding of the latter [\ref{Pfaffian and Supo}]).

Secondly, in the present paper we have given {\em examples} of rigid bundles. 
To be used most widely, the procedure should be generalised as most as possible.
By this we mean that, ideally, one starts with a (spectral cover) bundle $V$ on $X$ and
can provide a non-generic discrete twist class on the spectral cover surface $C$ such that
the twisted bundle $V'$ becomes rigid; thereby one would get for {\em each} spectral cover
bundle a closely related bundle $V'$ whose phenomenlogically relevant cohomological data 
usually have, from new discrete input parameters, 
even more flexibility than the original bundle $V$; thus for all phenomenological
relevant questions, as far as expressed in the cohomological data, 
one could work just as well with $V'$ instead of $V$, 
but one would have solved the many hard problems posed by the presence of the vector bundle moduli,
though not in the standard manner of ordinary moduli stabilisation 
but rather by some sort of rigidification or freezing.
To carry out steps in this direction one clearly has to generalize the investigations of
the present paper: after all the condition $h^1(End \, V)\!=\! CON_{\gamma}$ is, in explicit form
(cf.~(\ref{explicit rigidity condition})), a diophantine equation in variables 
with {\em integral} entries
(discrete bundle parameteres and similar parameters for $c\!=\!\pi_{C*}\chi$); so usually
one will get only some sort of reduction (corresponding to $CON_{\gamma}<h^1(End \, V)$) instead of 
a {\em complete} reduction. To solve this problem one would like to {\em combine}
the reduction effects of different twists.
But turning on a combined twist will demand only that the combined class exists giving
usually a less effective moduli reduction. But only in general. If one is in a set-up
where the existence of the sum of the classes implies already the existence of the individual classes
one {\em can} combine reduction effects (with the goal to reach a complete reduction). This will
be desribed elsewhere.

\noindent
I thank the DFG for support in the project CU 191/1-1 
and the FU Berlin for hospitality. 

\newpage

\appendix

\section{Appendix: Proof of proposition 2 (iii)}

\resetcounter

In this appendix we give the proof of proposition 2 $(iii)$ concerning the surjectivity of the 
map $(a_i)\lra a_0a_5^2-a_2a_3a_5+a_3^2a_4$ (the horizontal map in the 'upper way') used above.
For convenience of the reader we start to state again the assertion.

\noindent
{\bf \un{Proposition 2}} {\em (iii) The map
\beqa
\bigoplus_{i=0, \neq 1}^5 H^0\Big(B, \cO_B(\eta-ic_1)\Big)&\lra &
H^0\Big(B, \cO_B(3\eta-10c_1)\Big) ,
\eeqa
given by $(a_i)\lra a_0a_5^2-a_2a_3a_5+a_3^2a_4$, is surjective 
if $c_1$ and $\bar{\eta}$ are both big and nef.}

Note first that the assertion in question is, via a simple resacling of $a_2$,
trivially equi\-valent to the same assertion formulated for the expression
\beqa
\label{expression}
a_0a_5^2-2a_2a_3a_5+a_3^2a_4&=&\Delta_1 a_5 + \Delta_2 a_3
\eeqa
where we used the quantities
\beqa
\label{Delta expressions}
\Delta_1\; = \; a_0a_5-a_2a_3\; , \;\; \Delta_2 \; = \; a_3a_4-a_2a_5
\eeqa
These decompositions allow to adopt a two-step strategy to prove the surjectivity in question:
\begin{itemize}
\item in a step $(1a)$ and $(1b)$ one proves that the maps
from the space $\bigoplus_{i=0, \neq 1}^5 H^0\Big(B, \cO_B(\eta-ic_1)\Big)$ of the $(a_i)$ 
to the spaces $H^0(B, \cO_B(2\bar{\eta}+5c_1))$ of $\Delta_1$ and 
$H^0(B, \cO_B(2\bar{\eta}+3c_1))$ of $\Delta_2$, respectively, are surjective
\item in step $(2)$ one proves that the map from the spaces of the elements
$\Delta_1, a_5, \Delta_2, a_3$ to the space $H^0\Big(B, \cO_B(3\eta-10c_1)\Big)$
of the expression (\ref{expression}) is surjective.
\end{itemize}

Let us recall now Noether's $AF+BG$ Theorem: 
this concerns homogeneous polynomials $F$ and $G$ on ${\bf P^2}$ 
whose (zero-)divisors $(F)$ and $(G)$, i.e.~vanishing loci, 
do not have a component in common and intersect transversally; if $H$ is then a third
homogeneous polynomial and one has (with further homogeneous polynomials $A$ and $B$) a relation
\beqa
H&=&AF+BG
\eeqa 
one gets as necessary relation for such a decomposition the condition $(F)\cap (G)\subset (H)$;
the Theorem states that the latter condition is not only necessary but also already sufficient.

To use this in a more general context for various base surfaces $S$
let us introduce the following notation
\beqa
H&\in& H^0\Big(S, \cO(\delta)\Big)\\
A&\in& H^0\Big(S, \cO(\kappa)\Big)\\
F&\in& H^0\Big(S, \cO(\mu)\Big)\\
B&\in& H^0\Big(S, \cO(\lambda)\Big)\\
G&\in& H^0\Big(S, \cO(\nu)\Big)
\eeqa
such that one has the relation
\beqa
\delta&=& \kappa + \mu \; = \; \lambda + \nu
\eeqa
Crucial for the generalization of the proof of the Theorem from ${\bf P^2}$ to $S$
is the following quantity
\beqa
\rho:&=&\kappa - \nu \; = \; \lambda - \mu
\eeqa
The analogue of the Theorem will hold on $S$ if 
\beqa
\label{rho condition}
H^1\Big(S, \cO(\rho)\Big)&\stackrel{?}{=}&0
\eeqa
(cf.~the proof [\ref{GH}] of the standard form of the Theorem).
This holds if either $-\rho$ or $\rho+c_1$ is ample 
or at least (cf.~Th.~26, Ch.~1 [\ref{Fr}]) big and nef.
As in our application one has $\rho=3c_1$ and $2c_1$ in step $(1a)$ and $(1b)$, respectively,
and $\rho=\bar{\eta}+3c_1$ in step $(2)$ the condition (\ref{rho condition}) will
always be satisfied.\footnote{\label{ample footnote}Note that $\alpha$ ample and $\beta$ nef 
implies $\alpha+\beta$ ample, cf.~Exerc.~11, Ch.~1 [\ref{Fr}]; 
simlarly assuming $\alpha$ just to be big and nef and $\beta$ nef implies 
$\alpha+\beta$ big and nef as $\beta^2\geq 0$ (by Lemma 23, Ch.~1 [\ref{Fr}]) 
and also $\al \beta\geq 0$; 
the latter follows from $n\al\sim$ eff.~for $n\gg 0$ by Lemma 12, Ch.~1 [\ref{Fr}] 
(where $\al H >0$ for $H$ ample 
as again $nH\sim$ eff.~for $n\gg 0$ by Lemma 12, Ch.~1 [\ref{Fr}] 
such that $\al H\geq 0$ where $\al H\neq 0$ by $\al^2 >0$ and the Hodge index theorem)} 
This shows that we are done because one can easily see that the 
remaining conditions of the Theorem are also met as we explain now.

For example, in the \un{step (1a)} (the other cases are handled analogously) 
one asks whether an arbitrary element
\beqa
H&\in &H^0\Big(B, \cO_B(2\bar{\eta}+5c_1)\Big)
\eeqa
from the space where $\Delta_1$ lives
can be written as combination $H=AF+BG$ of some elements $F$ and $G$ 
(representing $a_5$ and $a_3$, respectively) with the help of further elements
$A$ and $B$ where one has (the minus signs in (\ref{Delta expressions}) do not matter, of course)
\beqa
A=a_0 &\in &H^0\Big(B, \cO_B(\bar{\eta}+5c_1)\Big)\\
F=a_5 &\in &H^0\Big(B, \cO_B(\bar{\eta})\Big)\\
B=a_2 &\in &H^0\Big(B, \cO_B(\bar{\eta}+3c_1)\Big)\\
G=a_3 &\in &H^0\Big(B, \cO_B(\bar{\eta}+2c_1)\Big)
\eeqa
Thus one has in this step 
\beqa
\delta  &=&2\bar{\eta}+5c_1\\ 
\kappa  &=& \bar{\eta}+5c_1\\
\mu     &=& \bar{\eta} \\
\lambda &=& \bar{\eta}+3c_1\\
\nu     &=& \bar{\eta}+2c_1
\eeqa 
giving indeed\footnote{if one would start instead from $\Delta_1=a_5a_0-a_3a_2$, say, 
one would get $\rho=-3c_1$ and similarly in all the other cases; 
given the sufficient conditions on $\rho$ for (\ref{rho condition}) to hold 
no difference results in the end}
the mentioned $\rho=3c_1$. 
We ask now whether for any such $H$ there exist $a_5$ and $a_3$ with
$(a_5)\cap (a_3)\subset (H)$ (where $(a_5)$ and $(a_3)$ should have no component in common
and should intersect transversally). Let us treat the problem in general:
these are $D\cdot D'$ conditions (where $D=(a_5)=(F)$ and $D'=(a_3)=(G)$ in our example)
while the degrees of freedom available for the effective divisors $D$ and $D'$ 
(note that here $H$ is {\em given}) are given, respectively, 
by\footnote{where we assume that $D+c_1$ and $D'+c_1$ are ample 
or at least (cf.~Th.~26, Ch.~1 [\ref{Fr}]) big and nef}
\beqa
h^0(B, \cO_B(D))-1&=&\frac{1}{2}D(D+c_1)\\
h^0(B, \cO_B(D'))-1&=&\frac{1}{2}D'(D'+c_1)
\eeqa
The number $m$ of the remaining degrees of freedom is 
$\frac{1}{2}\Big(D^2+(D+D')c_1+D'^2\Big)- D D'$ or 
\beqa
m&=&\frac{1}{2}\Big[ (D-D')^2+(D+D')c_1\Big]
\eeqa
Now $D$ and $D'$ are in any case effective, so the second term is nonnegative for $c_1$ being
ample or at least nef; the first term is strictly positive in our case of $((F)-(G))^2=4c_1^2$
giving in total a strictly positive dimension for the space of the remaining degrees of freedom.
Having elements $D=(F), D'=(G)$ with a {\em non}-transversal intersection 
would be a non-generic case (in a subspace of positive codimension), 
similarly elements with a common component; such degenerate cases can therefore be avoided.

\section*{References}
\begin{enumerate}

\vspace{-0.2cm}

\item
\label{FMW}
R. Friedman, J. Morgan and E. Witten, {\em Vector Bundles and F-Theory},
hep-th/9701162, Comm. Math. Phys. {\bf 187} (1997) 679.
 
\item
\label{C}
G.~Curio, {\em Chiral Matter and Transitions in Heterotic String Models},
hep-th/9803224, Phys.Lett. {\bf B435} (1998) 39.

\item
\label{C2}
G.~Curio, {\em Moduli restriction and Chiral Matter in Heterotic String Compactifications},
arXiv:1110.6315, JHEP01(2012)015.

\item
\label{DW}
Ron Donagi, Martijn Wijnholt,
{\em Higgs Bundles and UV Completion in F-Theory},
arXiv:0904.1218.

\item
\label{GH}
P.~Griffiths and J.~Harris ``{\em Principles of Algebraic Geometry}'' (1978) John Wiley \& Sons.

\item
\label{Fr}
R.~Friedman ``{\em Algebraic Surfaces and Holomorphic Vector Bundles} (1998) Springer. 

\item
\label{cxstrfix}
Lara B. Anderson, James Gray, Andre Lukas, Burt Ovrut,
{\em Stabilizing the Complex Structure in Heterotic Calabi-Yau Vacua},
hep-th arXiv:1010.0255, JHEP 1102:088,2011.\\
Lara B. Anderson, James Gray, Andre Lukas, Burt Ovrut,
{\em The Atiyah Class and Complex Structure Stabilization 
in Heterotic Calabi-Yau Compactifications},
arXiv:1107.5076.

\item
\label{Pfaffian and Supo}
Evgeny I. Buchbinder, Ron Donagi, Burt A. Ovrut,
{\em Superpotentials for Vector Bundle Moduli},
arXiv:hep-th/0205190, Nucl.Phys.B653:400-420,2003\\
Evgeny I. Buchbinder, Ron Donagi, Burt A. Ovrut,
{\em Vector Bundle Moduli Superpotentials in Heterotic Superstrings and M-Theory},
arXiv:hep-th/0206203, JHEP 0207 (2002) 066\\
Gottfried Curio, 
{\em On the Heterotic World-sheet Instanton Superpotential and its individual Contributions},
arXiv:1006.5568\\
Gottfried Curio,
{\em Perspectives on Pfaffians of Heterotic World-sheet Instantons},
arXiv:0904.2738, JHEP 0909:131,2009\\ 
Gottfried Curio,
{\em World-sheet Instanton Superpotentials in Heterotic String theory and their Moduli Dependence},
arXiv:0810.3087, JHEP 0909:125,2009.

\item
\label{het mod}
Gottfried Curio, Ron Y. Donagi,
{\em Moduli in N=1 heterotic/F-theory duality},
arXiv:hep-th/9801057, Nucl.Phys. B518 (1998) 603-631

\end{enumerate}
\end{document}